\documentclass[11pt]{article}
\usepackage{latexsym,amsthm,amsmath,amssymb,url}

\newtheorem{theorem}{Theorem}

\begin{document}

\title{Note on Perfect Forests}
\author{
Gregory Gutin\thanks{Royal Holloway, University of London, Egham,
Surrey, TW20 0EX, UK. This research was supported partially by Royal Society Wolfson Research Merit Award.}
}

\date{}

\maketitle

\begin{abstract}
\noindent A spanning subgraph $F$ of a graph $G$ is called {\em perfect} if $F$ is a forest,
the degree $d_F(x)$ of each vertex $x$ in $F$ is odd, and
each tree of $F$ is an induced subgraph of $G$.
We provide a short proof of the following theorem of A.D. Scott (Graphs \& Combin., 2001): a connected graph $G$ contains a perfect forest if and only if $G$ has an even number of vertices. 
\end{abstract}

\pagenumbering{arabic}
\pagestyle{plain}

\section{Introduction}\label{section:intro}
The number of vertices of a graph $G$ is called its {\em order}.
A spanning subgraph $F$ of a graph $G$ is called a {\em perfect forest} if 
\begin{itemize}
\item $F$ is a forest,
\item the degree $d_F(x)$ of each vertex $x$ in $V(F)$ is odd, and
\item each tree of $F$ is an induced subgraph of $G$.
\end{itemize}
Note that a perfect matching is a perfect forest. Clearly, if a connected graph $G$ has a perfect forest, then $G$ is of even order (as every graph with all vertices of odd degree must have an even number of vertices).
Scott \cite{Sco2001} proved that surprisingly the opposite implication is also true, i.e., every connected graph $G$ of even order contains a perfect forest.\footnote{In fact, Scott  \cite{Sco2001} formulated the result for arbitrary graphs, not just connected ones, with components of even order, but since the general result immediately follows from that on connected graphs, we restrict ourselves to connected graphs.}
Intuitively, it is clear that a perfect forest can provide a useful structure in a graph and, in particular, this notion was used by Sharan and Wigderson \cite{SW} to prove that the perfect matching problem for bipartite cubic graphs belongs to the complexity class ${\cal NC}$.
As a simple application of Scott's theorem, observe that a connected graph $G$ with even number $n$ of vertices, contains a spanning subgraph $H$ such that $|E(H)|\ge |E(G)|-2n+2$ and the parities of $d_H(x)$ and $d_G(x)$ are different for every vertex $x$ of $G$.

The proof of Scott's theorem in \cite{Sco2001} is graph-theoretical and relatively long consisting of two steps. In the first step, it is proved that the vertices of a connected graph $G$ of even order have a partition $V_1,\ldots ,V_t$ such that in each subgraph induced by $V_i$, all vertex degrees are odd. In the second step, it is shown that if such a partition cannot be refined further, then each induced subgraph is a tree. The proofs of both steps are constructive and can be turned into a polynomial algorithm to find a perfect forest.

In the next section, we provide a short proof of Scott's theorem using simple linear algebra arguments. Our proof can also be turned into a polynomial algorithm to find a perfect forest as there are polynomial algorithms to check linear independence of vectors and, if the vectors are linearly dependent, to find a nontrivial linear combination of them which is equal to the zero vector (one simply solves the corresponding system of linear equations). 

Related literature \cite{Caro,CarKraRod,RadSco,Sco1992} studies largest induced subgraphs with all odd degrees. The main conjecture on the topic mentioned in \cite{Caro} is that every graph $G$ of order $n$ without isolated vertices contains an induced subgraph $H$ of order $\Omega(n)$ such that all vertex degrees of $H$ are odd. To the best of our knowledge, while the conjecture was proved for trees \cite{CarKraRod,RadSco}, it remains open in general, with the best lower bound on the order of $H$ being $\Omega(n/\log n)$ \cite{Sco1992}. Linear-algebraic approach may be useful for the conjecture, too.

\section{Short Proof of Scott's Theorem}

\begin{theorem}\label{thmKcon1}
Let $G$ be a connected graph of even order $n$. Then $G$ 
contains a perfect forest. 
\end{theorem}
\begin{proof}
Let $V(G)=\{1,2,\ldots, n\}$ and let $K_n$ be the complete graph with vertex set $V(G)$. An edge $e\in E(K_n)$ with vertices $i$ and $j$ is denoted $ij$ and let $v(ij)$ be a vector in $\mathbb{F}_2^n$ in which the only nonzero coordinates are $i$ and $j$. Let $T$ be a spanning tree of $G$. Observe that vectors $\{v(f):\ f\in E(T)\}$ are linearly independent, and for every $e\in E(K_n)$, $v(e)$ is a linear combination of vectors in $\{v(f):\ f\in E(T)\}.$
Observe that all vertex degrees of a subgraph $H$ of $G$ are even if and only if $\sum_{e\in E(H)}v(e)=\mathbf{0},$ where $\mathbf{0}$ is the zero vector in $\mathbb{F}_2^n$, i.e., the vectors $\{v(e):\ e\in E(H)\}$ are linearly dependent.
 
Let $\mathbf{1}$ be the vector in $\mathbb{F}^n_2$ in which every coordinate is $1$. Observe that $\mathbf{1}$ is the sum of vectors $v(12),v(34),\ldots , v(n-1,n)$. Thus, $\mathbf{1}$ is the sum of vectors in $L\subseteq \{v(e):\ e\in E(T)\}.$
Observe that the vectors in $L$ are linearly independent. 
Let $M=\{v(e):\ e\in E(G)\}$, and for each $v(e) \in  M\backslash L$, let $S_{v(e)}=\{v(e)\}\cup L$. Consider two cases:

\begin{description}
\item[Case 1:] $S_{v(e)}$ is linearly dependent for some $v(e)\in M \backslash L$. Then we can find $L' \subseteq L$ such that $v(e)=\sum_{v\in L'} v$. Thus, we have a shorter expression for $\mathbf{1}$: $\mathbf{1} = v(e)+\sum_{v\in L\setminus L'}v$. Let $L:=\{v(e)\}\cup (L\setminus L')$ and note that vectors in $L$ are linearly independent.

  \item[Case 2:] $S_{v(e)}$ is linearly independent for each $v(e) \in M \backslash L$. 
 Consider the subgraph $F$ of $G$ induced by the edges $\{e\in E(G):\ v(e)\in L\}$. Observe that the degree $d_F(x)$ of each vertex $x\in V(G)$ is odd (as the sum of vectors in $L$ equals $\mathbf{1}$), $F$ has no cycle (as $L$ is linearly independent and thus all its subsets are linearly independent as well) and moreover if vertices $i$ and $j$ are in the same component of $F$, then $ij\not\in E(G)$ (as $S_{v(e)}$ is linearly independent for each $v(e) \in M \backslash L$).
  \end{description}

Since $s\le n$ and Case 1 produces a shorter expression for $\mathbf{1}$, after at most $n$ iterations of Case 1 we will arrive at Case 2.
\end{proof}

\end{document}